\documentstyle[11pt]{article}
\begin{document}

\renewcommand{\thefootnote}{\fnsymbol{footnote}}
\newcommand{\fracts}[2]{\textstyle \frac{#1}{#2} }
\newcommand{\fracds}[2]{\displaystyle \frac{#1}{#2} }
\newcommand{\eq}[1]{eq.(\ref{#1})}
%
\renewcommand{\theequation}{\arabic{section}.\arabic{equation}}
\newcommand{\sectione}[1]{ \section{#1} \setcounter{equation}{0} }
%
\newcommand{\startapp}{ \appendix
\renewcommand{\theequation}{\Alph{section}.\arabic{equation}} }
\newcommand{\bea}{\begin{eqnarray}}
\newcommand{\eea}{\end{eqnarray}}
\def\be{\begin{equation}}
\def\ee{\end{equation}}
\newcommand{\zerosection}{\setcounter{section}{-1}}

\def\sections#1#2{\bigskip\noindent{\bf #1\quad #2}\bigskip}
\def\proof{\medskip\noindent{\sl Proof\ } :\  }
\def\theorem#1{\medskip\noindent{\bf Theorem\ } #1 :\ }
\def\condition#1
{\medskip\noindent{\bf Condition\ } #1 :\ }
\def\corollary#1{\medskip\noindent{\bf Corollary\ } #1 :\ }
\def\proposition#1{\medskip\noindent{\bf Proposition\ } #1 :\ }
\def\definition#1{\medskip\noindent{\bf Definition\ } #1 :\ }
\def\question{\noindent{bf Question\ } :\ }
\def\qed{\rightline{\sl q.e.d.}\medskip}


\def\and{\quad{\rm and}\quad}
\def\OR{\quad{\rm or}\quad}
\def\for{\quad{\rm for}\quad}
\def\with{\quad{\rm with}\quad}
\def\Det{{\rm det}}
\def\Dirac#1{ #1 \hskip-6pt /\,}
\def\half{{1\over2}}
\def\p{\partial}
\def\Tr#1{{\rm T}\!{\rm r}\{#1\}}
\def\tr{{\rm T}\!{\rm r}}
\def\wed{\mathop{\wedge}}
\def\intdx#1{\int\! d^{#1}x\ }
\def\brs{\delta^{\ }_B{}}
\def\one{{\bf1}}

\def\hs{\hat{s}}
\def\R{{\widehat{\bf R}}}
\def\M{M{}}
\def\opst{{\bf p}}
\def\cp{{\cal P}}
\def\cpa#1{{\cal P}_A\big(#1\big)}

\def\al{{\cal A}}
\def\qal{{\cal M}}
\def\co{\Delta}
\def\col{\Delta_L}
\def\cor{\Delta_R}
\def\I{{\bf I}}
\def\tchi{\widetilde{\chi}}
\def\e{{\varepsilon}}
\def\A{{\bf A}}
\def\tA{\widetilde{\bf A}}
\def\btheta{\bar{\theta}}
\def\C{{\rm C}}
\def\X{{\bf X}}
\def\d{{\bf d}}
\def\norm{{\cal N}}
\def\antipode#1{\kappa\big(#1\big)}
\def\ap{S}
\def\bra{[\![}
\def\ket{]\!]}
\def\anti{\kappa}
\def\cs{{\cal S}}

\zerosection

\begin{titlepage}
\vspace*{-4ex}


\null \hfill HD-THEP-92-39  \\
\null \hfill Preprint TU-411  \\
\null \hfill September, 1992 \\[4ex]

\begin{center}

\bigskip
\bigskip

{\Large \bf
Quantum Deformation of BRST Algebra${}^{{}^\ddagger}$
}
\footnotetext[3]{This work is partly supported by Alexsander von Humboldt
Foundation.}
\\ [18ex]


  Satoshi \, Watamura\\[5ex]

Department of Physics, College of General
Education, \\[1ex]
Tohoku University, Kawauchi, \\[1ex]
Aobaku, Sendai 980, Japan.\\[1ex]
bitnet:watamura@jpntuvm0\\[4ex]

Institute f\"ur Theoretische Physik\\[1ex]
Universit\"at Heidelberg\\[1ex]
Philosophenweg 16, D-6900\\[10ex]

%
 \end{center}

\bigskip
\begin{abstract}
\medskip
We investigate the $q$-deformation of the
BRST algebra, the algebra of the ghost,
matter and gauge fields on one spacetime point using the result of the
bicovariant
differential calculus.  There are two nilpotent operations in the algebra, the
BRST transformation $\brs$ and the derivative $d$.  We show that one can define
the covariant commutation relations among the fields and their derivatives
consistently with these two operation as well as the $*$-operation, the
antimultiplicative inner involution.
 \end{abstract}
 \end{titlepage}

\section{Introduction}
\vskip 0.3cm

It is an interesting question whether one can construct a $q$-analogue
of the gauge theory by taking the quantum group [Dri,FRT,Jim,Wor1]
as a symmetry.
One of the interesting possibilities of such a $q$-deformed theory is that the
deformation parameter $q$ may play a role of a regularization parameter.
Furthermore, since the quantum group is provided by a noncommutative algebra,
in such a theory the noncommutative geometry [Connes]
plays a basic role like the
differential geometry in the usual gauge theory.

There are some proposals to this problem [Are,Ber,BM,Hira,IP,WuZ].  However,
it seems that there
are still conceptual problems concerning the definition of the gauge
transformation when we take the quantum group as an algebraic object of the
gauge symmetry.
Since the quantum group is formulated in the language of the Hopf algebra, it
forces us to formulate the whole theory in algebraic language.  Thus the
gauge transformation will be represented in an abstract language and the term
of the transformation parameter becomes obscure. Even when we consider only the
infinitesimal transformation, we have still the question of the definition of
the infinitesimal parameters.

One of the alternative formulations of the gauge theory is given by the BRST
formalism [BRST,KO].
There, the gauge transformation parameter is replaced by the ghost
fields and becomes an object of equal level with the matter and the gauge
fields.
Therefore, when we consider the $q$-deformation of the gauge theory, it is very
natural to consider the $q$-deformed field algebra starting with the BRST
formalism.

In this paper we construct a $q$-deformation of the BRST algebra which is the
algebra of the gauge fields, the ghost fields and appropriate matter fields on
one spacetime point.  The gauge transformation of the theory is replaced by the
BRST transformation which is represented by a nilpotent "differential operator"
$\delta_B$.

The paper is organized as follows.  In section 1, we give a collection of
some results of the bicovariant differential calculus on the
quantum group which we
need for the later investigation. In section 2, we discuss about the algebraic
properties of the gauge transformation and we give the general conditions that
the $q$-deformed BRST algebra must satisfy. Following this general framework,
we define the algebra and prove various consistencies in section 3.
Section 4 is devoted to discussions.

For the notation concerning the Hopf algebra [Abe], we take:
the coproduct $\co$, the antipode $\anti$ and the counit $\e$ (see also
ref.[CW] for our notation).
Through this paper the upper case roman index $I,J,K,L$ runs $0,-,3,+$ and the
lower case index like $a,b,c,d$ runs over the label of the adjoint
representation, $-,3,+$, otherwise we specify explicitly.

\sectione{Bicovariant Differential Calculus}
\vskip 0.3cm

Before we start to construct the BRST algebra, let us briefly recall some
results of the bicovariant differential calculus [Wor2,Rosso,Jur,CSWW].
The one-forms are defined by the right invariant bases $\theta^i_j$ ($i,j=1,2$)
where $\theta^i_j{}^*=\theta^j_i$.  Using the spinor metric
\be%
\epsilon^{kl}
= \left( \matrix{ 0&-q^{-\half} \cr
		q^{\half}&0 } \right)
\quad,
\label{a0}
\ee%
we define $\theta^{ij}=\theta^i_k\epsilon^{kj}$, then they have the commutation
relation
\be%
a\theta^{ij}=\theta^{kl}(a*{\bf L}^{ij}_{kl})
\label{a1}
\ee
for $\forall a\in Fun_q(SU(2))$.

${\bf L}$ is the functional $Fun_q(SU(2))\rightarrow \C$ defined by
\be%
{\bf L}^{ij}_{kl}=(L_-{}^j_l*L_+{}^i_k)\circ\anti
\quad,
\label{a2}
\ee%
where the functionals
$L_{\pm}$ and the convolution product are defined in
the appendix A.
  [The functionals $f_{\pm}$ appearing in ref.[CSWW]
equivalent to the $L_{\pm}$ in ref.[FRT] which we use here. Thus, ${\bf
L}^{ij}_{kl}$ is equivalent to the functional $f^{ij}_{{\rm Ad}\,kl}\circ\anti$
in ref.[CSWW].]

The right invariant bases $\theta^{ij}$ can be split into two part as a left
comodule : the adjoint representation $\theta^a$ ($a=-,3,+$) and the singlet
$\theta^0$.
We use the $q$-Pauli matrices $\sigma^I_{ij}$ and $\sigma_I^{ij}$ where
$\sigma^I_{ij}\sigma_J^{ij}=\delta^I_J$ ($I=0,-,3,+$) and
\be%
\sigma^0_{kl}={q \over \sqrt{Q}}\epsilon_{kl},
\quad
\sigma^+_{kl}=\left( \matrix{ 0&0 \cr
		0&-1 } \right)
,\quad
\sigma^-_{kl}=\left( \matrix{ 1&0 \cr
		0&0 } \right)
,\quad
\sigma^3_{kl}={-1 \over \sqrt{Q}}\left( \matrix{ 0&q^{\half} \cr
		q^{-\half}&0 } \right)
,\label{a5}
\ee%
where $Q=q+q^{-1}$ and $\epsilon_{kl}=-\epsilon^{kl}$. With these $q$-Pauli
matrices we can write the projectors as
\bea%
{\cal P}^{ij}_{A\,kl} &=&\sigma^{ij}_0\sigma^0_{kl}\quad,\label{a5.1}\\
{\cal P}^{ij}_{S\,kl} &=&\sigma^{ij}_a\sigma^a_{kl}\quad,\label{a5.2}
\eea%
where ${\cal P}_S$ ( ${\cal P}_A$) is the projector for the
$q$-(anti)symmetric product.
Using these $q$-Pauli matrices we define
\be%
\theta^I\equiv\sigma^I_{ij}\theta^{ij}
\quad.
\label{a5.3}
\ee%

The $q$-deformed exterior derivative ${\bf d}$ is defined as a map from
$Fun_q(G)$ to the bimodule defined with the basis $\theta$ requiring that the
nilpotency and the Leibniz rule hold in the standard way [Wor2].
Such an operation
 can be defined simply as the commutator with the singlet component $\theta^0$
 as [CSWW]
\be%
{\bf d}a=
{ig\over \omega}[\theta^0,a]_-
\label{a6}
\ee%
for any element $a\in Fun_q(SU(2))$, where $\omega=q-q^{-1}$, $i=\sqrt{-1}$ and
$g$ is a non-zero real constant. [The relation of the constant $g$ with
the constant $N_0$ in ref.[CSWW] is
 $g={-i\sqrt{Q}\over q N_0}. $]
Since ${\bf d}a$  is an element of the bicovariant bimodule, we can expand it
with the basis as
\be%
{\bf d}a=
\theta^I(a*\chi_I)
\quad,
\label{a6.1}
\ee%
where the right invariant vector field $\chi_{ij}$ is given by
\be%
\chi_I\equiv\sigma_I^{ij}\chi_{ij}=\sigma_I^{ij}
{ig\over\omega}(\sigma^0_{ij}\e-\sigma^0_{kl}{\bf L}^{kl}_{ij})
\quad.
\label{a7}
\ee%
 We can consider the functional $\chi_I$ as a differential operator. The
 Leibniz rule is given by the coproduct of $\chi_I$:
\bea%
(ab*\chi_I) &=&(a\otimes b)*\co(\chi_I)\quad,\nonumber\\
 &=&(a*\chi_I)b+(a*{\bf L}^J_I)(b*\chi_J)\quad.\label{a7.1}
\eea%

One of the suggesting relation given by the bicovariant differential calculus
is the $q$-analogue of the Maurer-Cartan equation.
In ref.[CSWW], we gave the expression in a more familiar form:
\bea%
{\bf d}\theta^0&=&0\quad,\label{a8}\\
{\bf d}\theta^a&=&
{-ig\over q^2+q^{-2}}f^a_{bc}\theta^b\wedge\theta^c
\quad,
\label{a9}
\eea%
where $\wedge$ is the $q$-deformed exterior product.   The $f^a_{bc}$ is the
$q$-analogue of the structure constants.  Using the general formula for the
structure constants in ref.[CSWW] (See also ref.[Carow]), we obtain them for
the $Fun_q(SU(2))$ as
\bea%
f^+_{+3} &=&q\quad f^+_{3+}=-q^{-1}\quad,\nonumber\\
f^-_{3-} &=&q \quad f^-_{-3}=-q^{-1}\quad,\nonumber\\
f^3_{-+} &=&1\quad f^3_{+-}=-1\quad f^3_{33}=q-q^{-1}
\quad.
\label{a10}
\eea%

The Maurer-Cartan equation and the definition of the $\chi_I$ given in \eq{a7},
we can deduce the commutation relation among the $q$-vector fields $\chi_I$.
We find that $\chi_0$ is central and actually proportional to the second
Casimir operator [Weich,CSWW].
The commutation relations of others are given by
\be%
{\bf P}\!_{Ad\, bc}^{\ b'c'}(\chi_{b'}*\chi_{c'})
={-1 \over q^2+q^{-2}}f^a_{bc}\opst*\chi_a
\quad,
\label{a11}
\ee%
where the functional $\opst$ is central and
\be%
\opst=
ig\e-\omega\chi_{0}
\quad.
\label{a12}
\ee%
(See also eq.(5.50) in ref.[CW].) The matrix
${\bf P}\!_{Ad}=({\cal P}_S,{\cal P}_A)+({\cal P}_A,{\cal P}_S)$ is an operator
projecting the tensor onto the $q$-antisymmetric part, where
$({\cal P}_r,{\cal P}_{r'})$ is defined in eq.(7.4) of ref.[CSWW]
(see appendix B).
${\bf P}^{\ ab}_{Ad\, cd}$ is the projector restricted to the
product of two adjoint representations.
Thus l.h.s. of \eq{a11} gives the $q$-analogue of the usual commutator of the
generators.

For the $Fun_q(SU(2))$ calculus we can write
\be%
{\bf P}\!_{Ad\ cd}^{\ ab}={-1 \over q^2+q^{-2}}f^{a'}_{cd}f_{a'}^{ab}
\quad,
\label{a12.1}
\ee%
where $f_a^{bc}\equiv -f^a_{bc}$.
 Therefore, it is straightforward to evaluate these projection operators and
 using that result, \eq{a11} is written as
\bea%
q^{-1}\chi_3*\chi_+-q\chi_+*\chi_3&=&\opst*\chi_+\quad,\label{a13}\\
q\chi_3*\chi_--q^{-1}\chi_-*\chi_3&=&-\opst*\chi_-\quad,\label{a14}\\
\chi_+*\chi_--\chi_-*\chi_+-\omega\chi_3*\chi_3 &=&\opst*\chi_3
\quad.
\label{a14'}
\eea%

In the limit $q\rightarrow1$, the operator $\opst$ is proportional to the
counit and we get the standard commutation relation of generators of $SU(2)$.

\sectione{Gauge Transformation and BRST Formalism}

\subsection{Gauge Transformation}

As we explained in the introduction, we need to represent the gauge theory
using an appropriate algebra languages which fits
to the Hopf algebra structure.
Thus, let us first reconsider the gauge and the BRST transformation in the
non-deformed theory.  We take $SU(2)$ gauge theory as an example but the result
applies to the general group.

When we consider the usual non-deformed gauge theory with a symmetry group
$SU(2)$, the matter like
a lepton is represented by the field which is the section of the associated
fiber bundle of the structure group $SU(2)$ with the spacetime as a base
manifold $B$.  Thus the algebra of the matter fields is the algebra
of all possible sections.

Under the gauge transformation, the matter field $\Psi$ is
transformed according to its representation.  Giving the $SU(2)$ valued
function $g(x)\in SU(2)$ on the base manifold $B\ni x$, when the matter is
 of the fundamental representation
 the gauge transformation of the matter $\Psi^i(x)$ can be written as
\be%
[\Psi^{i}(x)]^g=M^i_j(g(x))\Psi^j(x)
\quad,
\label{ba}
\ee%
where ($i,j=1,2$). We wrote the gauge transformation matrix as $M^i_j(g(x))$ to
clarify the algebraic structure.  The matrix element $M^i_j$ maps the $g(x)$ to
the complex valued function on the base manifold and thus pointwise $M^i_j$ is
an element of the $Fun(SU(2))$.
Therefore, the gauge transformation property of the matter field can be
translated into the algebraic language such that
the algebra of matter fields is the (left)comodule algebra, and
there is a pointwise (left)coaction $\col$ of $Fun(SU(2))$ on the field
$\Psi$:
\be%
\col(\Psi)=\sum_s T_s\otimes \Psi^s
\quad,
\label{b2}
\ee%
where $T_s\in Fun(SU(2))$ are matrix elements of the representation
corresponding to the matter $\Psi$. For the fundamental representation \eq{b2}
is $\col(\Psi^i)=M^i_j\otimes\Psi^j$ and with the corresponding argument we get
\eq{ba}.

\medskip

The infinitesimal transformation corresponding to the transformation (\ref{ba})
can be written as
\be%
\delta_\xi(\Psi^i(x))=\xi^a(x)\chi_a(M^i_j)\Psi^j(x)
\quad,
\label{b5}
\ee%
where $a=-,3,+$ is the label of the adjoint representation of $SU(2)$,
$\xi^a$ is the gauge parameter which is the real function of the spacetime and
$\chi_a(M^i_j)$ is a $2\times 2$ matrix.
In the non-deformed
case we can identify $\chi_a$ with the right invariant vector
fields which are considered as the linear functionals
$Fun(SU(2))\rightarrow\C$ with the evaluation
\be%
\chi_a(M^i_j)=L_a^\mu{\p\over\p\phi^\mu} M^i_j(g(\phi^\nu))\Big|_{at
\ unity}
\quad,
\label{b6}
\ee%
where $g(\phi^\mu)$ is the group element parametrized by $\phi^\mu$ and
$L^\mu_a$ is the component of the right invariant vector field. The r.h.s.
gives the Pauli matrix for the $SU(2)$ case and thus \eq{b5} is the familiar
infinitesimal transformation.
The above structure can be translated into the algebraic language as follows:

The infinitesimal transformation $\delta_\xi$ of the matter field $\Psi$ can be
represented by the vector fields $\chi_a$ and the infinitesimal parameter
$\xi^a$ as
\be%
\delta_\xi\Psi=\xi^a(\Psi*\chi_a)
\quad,
\label{b3}
\ee%
where $(\cdot *\cdot)$ denotes the convolution product of a comodule with a
functional. For detail see appendix A.
For the fundamental representation, using the definition (\ref{bb}) it is easy
to show that the formula (\ref{b3}) is equivalent to \eq{b5}.

\medskip

Using the above algebraic representations,
we may consider the $q$-analogue of the
finite and the infinitesimal gauge transformation which we will discuss
elsewhere.  Here we want to concentrate on the $q$-deformation of the BRST
algebra which seems the most
appropriate algebra to consider the $q$-deformation of
the gauge theory.

The BRST transformation of the matter field is defined by replacing the gauge
parameter $\xi^a$ by the ghost field $C^a$ [FP].
Thus the BRST transformation can be written as
\be%
\delta_B\Psi=C^a(\Psi*\chi_a)
\quad.
\label{b7}
\ee%
For the fundamental representation this is
\be%
\delta_B\Psi^i=C^a\chi_a(M^i_j)\Psi^j
\quad.
\label{b8}
\ee%
Replacing the $\chi_a(M^i_j)$ with the Pauli matrix this is a familiar BRST
transformation.

\subsection{Definition of $q$-deformed BRST Algebra}

After the above preparation we specify the properties of the $q$-deformed BRST
algebra which we will construct in the next section.  We extract appropriate
properties from the non-deformed BRST formalism and impose them as the
condition which our algebra should satisfy.  We also require that in general
 in the limit $q\rightarrow1$, we always get the algebra equivalent to  the
non-deformed one.

The BRST algebra is the algebra which contains the matter fields $\Psi$ and the
gauge fields $A^I$
and the ghosts $C^I$ which are the standard field contents of
the BRST formalism. The suffix $I$ corresponds to the adjoint representation in
non-deformed case.  However in the $q$-deformed case we only require that it
contains the adjoint representation and allow to add a singlet component like
the right invariant basis
 $\theta^i_j$ in the bicovariant differential calculus.

In the field theory, we have the spacetime derivative $d$ and therefore, we
also require the existence of the map $d$ in the algebra which maps:
\be%
(\Psi,\  A^I,\  C^I)\quad {\buildrel d\over\longrightarrow}
\quad  (d\Psi,\ dA^I,\ dC^I) \quad{\buildrel d\over\longrightarrow}\quad 0.
\label{bx1}
\ee%
To construct the algebra we treat
the fields $d\Psi$, $dA^I$ and $dC^I$ as independent generators
from the original fields and then require the consistency with the above map
$d$.

\definition{1}

{\sl
The BRST algebra $\al_B$ is a comodule algebra over $Fun_q(G)$ which is
generated by the following set of the comodules:
\be%
\al_B=\C<C^I,\Psi,A^I,dC^I,d\Psi,dA^I>/{\cal I}
\quad,
\label{bx2}
\ee%
where $C^I$ represents the ghost, $\Psi$ the matter and $A^I$ the
gauge fields.  ${\cal I}$ is a set of the covariant commutation relations among
these comodules, which we shall determine in the next section.
}
\medskip

In the non-deformed BRST formalism of the gauge theory there are two nilpotent
operations, the exterior derivative $d$ and the BRST transformation $\brs$.
 We also require the corresponding structure in the algebra and
 that they keep
 the following properties in the $q$-deformed case:

\condition{A}

{\sl The BRST algebra possesses the following operations:
\medskip
\begin{enumerate}
\item
There exists an operation $\brs$ in the algebra $\al_B$ such that
\begin{enumerate}
\item{the $\brs$ operation satisfies the Leibniz rule in the graded sense,}
\item{ $\brs^2=0$.}
\end{enumerate}
\item
There exists an operation $d$ corresponding to the exterior derivative  such
that\begin{enumerate}
\item{ the $d$ operation satisfies the Leibniz rule in the graded sense,}
\item{ $d^2=0$,}
\item{ the action on $A^I$, $\Psi$ and $C^I$ is defined as \eq{bx1}.}
\end{enumerate}
\item{The two operations are anticommuting: $d\brs+\brs d=0$.}
\item
There exists a $*$-conjugation which is an inner involution of the algebra
$\al_B$ and antimultiplicative in the graded sense and also satisfies
\begin{enumerate}
\item{ \quad$\brs\circ {}^*={}^*\circ\brs$,}
\item{ \quad$d\circ {}^*={}^*\circ d$.}
\end{enumerate}
\item
The operations $\brs$ and $d$ are covariant:
For any element $\rho\in\al_B$ they satisfy
\begin{enumerate}
\item{$ \col(\brs\rho)=(id\otimes \brs)\col(\rho)$,}
\item{$ \col(d\rho)=(id\otimes d)\col(\rho)$.}
\end{enumerate}
\end{enumerate}
}
\bigskip

In the non-deformed case the number of the ghost fields is the dimension of the
adjoint representation, i.e., $3$ for $SU(2)$.
It is known that the property of the ghosts under the BRST transformation is
related with the invariant one-forms on the group.
In the $q$-deformed case, the result of the bicovariant bimodule calculus
implies
that the
number of the independent bases of the invariant one-forms is $4$ for the
calculus on $Fun_q(SU(2))$.  They include both the adjoint and singlet
representation.
Although the ghost fields are not required to be the bimodule over
$Fun_q(SU(2))$, it turns out that when we consider the covariant commutation
relation with the different type of fields such as the matter fields, the
projection onto the adjoint components are not compatible with the commutation
relations.
Therefore, in the $q$-deformed BRST algebra, we introduce the four ghosts $C^I$
where the suffix $I$ runs $0,-,3,+$.

For the convenience for the following discussion, first we give here the
definition of the ghosts $C^I$:

\definition{2}

{\sl
In the $q$-deformed BRST algebra based on the bicovariant differential calculus
on $Fun_q(SU(2))$, we define the ghost field as a comodule represented by a
 $2\times2$ matrix
$C^i{}_j$.  The left-coaction on it is
\be%
\co_L(C^i{}_j)=M^i_{i'}\anti(M^{j'}_j)\otimes C^{i'}{}_{j'}\quad,
\quad,
\label{c01}
\ee%
and under the $*$-conjugation it transforms like a hermitian matrix:
\be%
  (C^i{}_j)^*=C^j{}_i
\quad.
\label{c02}
\ee%
}
\medskip

We also introduce the upper index object $C^{ij}$ for convenience by using the
spinor metric $\epsilon^{ij}$:
\be%
C^{ij}=C^{i}{}_{k}\epsilon^{kj}
\quad,
\label{c03}
\ee%
with this basis the coaction becomes simply
\be%
\co_L(C^{ij})=M^i_{i'}M^j_{j'}\otimes C^{i'j'}
\quad.
\label{c04}
\ee%

We can decompose the ghost fields into singlet and adjoint representation by
using the $q$-Pauli matrix given in \eq{a5} as
\be%
C^I=\sigma^I_{i_1i_2}C^{i_1i_2}
\quad,
\label{c05}
\ee%
where $I=0,-,3,+$.
The singlet component $C^0$ is invariant under the left-coaction.

Note that like \eq{c05}, in the following we freely change the suffix $I,J,..$
with double suffix $\{i_1i_2\},\{j_1j_2\}...$ by using the $\sigma$ matrix.

The ghosts are the anticommuting fields in the BRST formalism, and therefore,
to
define the algebra of the ghosts $C^I$, we also impose the $q$-deformed
anticommutativity among the ghosts. For this purpose, we take the same
definition for the product rule as the one of the $\wedge$ product in the
bicovariant differential calculus.

Finally, in the non-deformed case, the BRST transformation of the ghosts has
the same form as the Maurer-Cartan equation.  Therefore for the $q$-deformed
BRST algebra, we postulate that the BRST transformation of the ghost fields has
the same form as the Maurer-Cartan equation of the bicovariant differential
calculus (\ref{a8}) and (\ref{a9}).

Here we summarize the conditions on the $q$-deformed ghosts:

\condition{B}

{\sl
\begin{enumerate}
\item
As a comodule, they have the same properties as the right invariant basis
$\theta^i_j$ appearing in the bicovariant differential calculus and consist of
both adjoint and singlet components.
\item
They are $q$-anticommuting.
\item
The BRST transformation $\brs$ of the ghosts has the same form as the
Maurer-Cartan equation obtained by the bicovariant differential calculus.
\bea%
\brs C^a
 &=&{-ig\over q^2+q^{-2}}f^a_{bc}C^bC^c
 \quad,\label{bx3}\\
\brs C^0 &=&0\quad,\label{bx4}
\eea%
where $g$ is an arbitrary non-zero real constant.  The structure constant
$f^a_{bc}$ for $Fun_q(SU(2))$ is given in \eq{a10}.
\end{enumerate}
}
\medskip

To define the $q$-deformed BRST transformation of the matter, we take the
algebraic representation in \eq{b7}.  Therefore, using the above ghost fields
we define the $q$-deformed BRST transformation of the matter analogously to
\eq{b7} as
\be%
\brs \Psi= C^I(\Psi*\chi_I)=C^a(\Psi*\chi_a)+C^0(\Psi*\chi_0)
\quad,
\label{b9}
\ee%
where $\chi_I\in {\cal U}_q(SU(2))$ is the one given in \eq{a7}. Although the
last term does not have the corresponding term in the non-deformed case, it
goes to zero in the limit $q\rightarrow1$.
The singlet component of the ghost is not desirable from the physical point of
view. On the other hand, as we shall see it seems it is necessary to include it
 in order to put the algebra
 into a simple form.  We come back to this point in the discussion.

\medskip
Therefore, for the $q$-deformed BRST transformation of the matter we have the
following condition:

\condition{C}

{\sl
The BRST transformation
 of the matter fields is defined by the $q$-analogue of the
infinitesimal transformation with the ghost as the parameter:
\be%
\brs \Psi=C^I(\Psi*\chi_I)
\quad,
\label{bx5}
\ee%
where the $\chi_I$ are the functionals given in \eq{a7}.
}
\medskip

Finally we require the existence of the covariant derivative which is
represented by the derivative $d$ and the gauge fields $A^I$.  The coupling of
the gauge fields to the matter
fields is determined naturally by the structure of
the BRST transformation of the matter fields given in \eq{bx5}.  Therefore, our
requirement concerning the covariant derivative is:

\condition{D}

{\sl
There exists a covariant derivative $\nabla$ which acts on the matter as
\be%
\nabla\Psi=d\Psi+A^I(\Psi*\chi_I)
\quad,
\label{bx6}
\ee%
where $A^I$ are the gauge fields which satisfies
\be%
(A^i_j)^*=A^j_i
\quad.
\label{bx6.1}
\ee%
The covariant derivative transforms with the same rule as the corresponding
matter
\be%
\brs\nabla\Psi=C^I(\nabla\Psi*\chi_I)
\quad.
\label{bx7}
\ee%
Under the $*$-conjugation it has the property
\be%
\nabla\circ{}^*={}^*\circ\nabla
\quad.
\label{bx8}
\ee%
}
\medskip

In the following sections, requiring the above conditions and the covariance,
we define the comodule algebra $\al_B$.  The main part of the construction is
to define the commutation relations ${\cal I}$.  The relation ${\cal I}$ is
defined by the following requirement.

\condition{E}

{\sl
\begin{enumerate}
\item{The covariance, i.e. if the relation $r=0$ then $\col(r)=0$.}
\item{The consistency of the relations among each others.}
\item{Invariance under $*$-conjugation.}
\end{enumerate}
}
\medskip

\sectione{The BRST Algebra}

In this section we give the commutation among the elements and the BRST
transformation of the gauge fields to complete the definition of the BRST
algebra.  For convenience, we give all relations in the first part of this
section.  The proof of the consistency of these relations are collected
in the remaining part of the section.

\subsection{Results}

The commutation relation of each type
of fields among themselves can be defined by taking
the $q$-antisymmetric ($q$-symmetric) product to vanish if it is
a bosonic(fermionic) field in the limit $q\rightarrow1$.

The ghost fields are
$q$-anticommuting by definition.  The gauge fields are
also $q$-anticommuting since they are spacetime one-forms
in the limit of $q\rightarrow1$.
We define the $q$-anticommutation relation of these fields using the same
formula used to define the $\wedge$ product in ref.[CSWW]:
\bea%
({\cal P}_S,{\cal P}_S)^{IJ}_{KL}C^KC^L &=&0\quad, \label{c1}\\
({\cal P}_A,{\cal P}_A)^{IJ}_{KL}C^KC^L &=&0\quad, \label{c2}
\eea%
\bea%
({\cal P}_S,{\cal P}_S)^{IJ}_{KL}A^KA^L &=&0\quad, \label{cc1}\\
({\cal P}_A,{\cal P}_A)^{IJ}_{KL}A^KA^L &=&0\quad. \label{cc2}
\eea%
For the notation $(\cdot,\cdot)$ for the pair of projectors see appendix B.

The other relations including the derivative of the fields have also to be
defined. They must satisfy the consistency condition {\bf E}.  Furthermore,
since the
operation $d$ relates some of the relations, they are not
 all independent.  The independent
relations are the ones between $(\{C^I\},\{dC^I\})$, $(\{A^I\},\{dA^I\})$,
$(\{C^I\},\{A^I\})$, $(\{C^I\},\{\Psi\})$, $(\{A^I\},\{\Psi\})$,
$(\{\Psi\},\{\Psi\})$ and $(\{\Psi\},\{d\Psi\})$.
The other relations can be
derived from them by requiring the consistency with the $d$ operation.

When we require the consistency with other structures, we can also fix those
relations. The resulting relations except the $(\{\Psi\},\{\Psi\})$ and
the $(\{\Psi\},\{d\Psi\})$ relations are given by the following:

\proposition{1}

{\sl
Define the ordering of the fields as
\be%
\{\Psi,d\Psi\}>\{dA^I\}>\{A^I\}>\{dC^I\}>\{C^I\}
\quad,
\label{cc3}
\ee%
then if $X>Y^I$, the commutation relation is given by
\be%
X Y^I=\pm Y^J(X*{\bf L}^I_J)
\quad,
\label{cc4}
\ee%
where the sign is taken as $+$($-$) if they are commuting(anticommuting) in the
limit $q\rightarrow1$ and ${\bf L}^I_J$ is the functional defined in \eq{a2}.
 Note that we take the $1$-form and the ghost
anticommuting with each other.

The relations of $dA^I$ and $dC^I$ are
\bea%
({\cal P}_S,{\cal P}_A)^{IJ}_{KL}dA^KdA^L&=&0\quad,\label{cc5}\\
({\cal P}_A,{\cal P}_S)^{IJ}_{KL}dA^KdA^L&=&0\quad,\label{cc6}
\eea%
and
\bea%
({\cal P}_S,{\cal P}_A)^{IJ}_{KL}dC^KdC^L&=&0\quad,\label{cc7}\\
({\cal P}_A,{\cal P}_S)^{IJ}_{KL}dC^KdC^L&=&0\quad,\label{cc8}
\eea%
}

The last two relations simply mean that $dA^I$ and $dC^I$ are $q$-commuting
fields as expected.

The algebra of the matter fields can be defined like a quantum plane, since
the quantum plane algebra is the algebra generated by the comodule imposing an
appropriate commutation relation [Manin].
The algebra depends on the representation of the matter fields in the model.
In our construction, we do not need to specify the representations of the
matter.  The algebra of the ghost and gauge fields which is defined in this
section is applicable for any representation of the matter.  This property
provides the flexibility to consider the model with various matter fields.
We give one example in the appendix C.

With the above relations
we can find the BRST transformation of the gauge field by
using the standard logic to define it in the field theory.

\proposition{2}

{\sl
The BRST transformation of the gauge field is given by
\bea%
\brs A^0 &=&dC^0\quad,\label{cc9}\\
\brs A^a&=&dC^a-
ig(\omega C^0A^a+f^a_{bc}C^bA^c)\quad,\label{cc10}
\eea%
and it is nilpotent.
}

This completes the definition of the algebra.  In the rest of this section we
give the proof of the consistency of the
above relations and the nilpotency of the
BRST operation.

In the last part of this section we also define the field strength using
the above algebra.
The result is

\proposition{3}

{\sl
The field strength is given by
\bea%
F^a
&=& dA^a-{ig \over q^2+q^{-2}}f^a_{bc}A^bA^c
\quad,\label{cc11}\\
F^0&=&dA^0\quad.
\label{cc12}
\eea%
The field strength is covariant under the BRST transformation:
\be%
\brs F^I=C^J(F^I*\chi_J)
\quad,
\label{cc13}
\ee%
and satisfies the Bianchi identity:
\bea%
d F^a&=&{ig\over q^2+q^{-2}} f^a_{bc}[A^bF^c-F^bA^c]
\quad,
\label{cc14}\\
dF^0&=&0\quad.\\
\eea%
}

\subsection{Algebra of Ghosts}

Here we prove the consistency of the commutation relation of the ghosts
\eq{c1} and \eq{c2}.

{}From the definitions (\ref{c1}) and (\ref{c2}),
 we obtain for the component of the adjoint representation:
\bea%
C^+ C^+ &=&0\label{c3}\quad,\\
C^-C^- &=&0\label{c4}\quad,\\
C^3C^3 &=&\omega C^-C^+\quad,\label{c5}\\
C^+C^3 &=&-q^{2}C^3C^+\quad,\label{c6}\\
C^-C^3 &=&-q^{-2}C^3C^-\quad,\label{c7}\\
C^+C^- &=&-C^-C^+ \quad.\label{c8}
\eea%
The relations including the singlet component are
\be%
[C^0,C^a]_+={-\omega\over q^2+q^{-2}} f^a_{bc}C^bC^c\quad,
\label{c9}
\ee%
or explicitly,
\bea%
[C^+,C^0]_+&=&{q\omega }C^3C^+\quad,\label{c10}\\ [0ex]
[C^-,C^0]_+&=&{q\omega}C^-C^3\quad,\label{c11}\\ [0ex]
[C^3,C^0]_+&=&\omega C^+C^-\quad.\label{c12}
\eea%
Finally we have
\be%
C^0C^0=0\quad.
\label{c13}
\ee%

Since the commutation relations among the ghosts are the same as the right
invariant basis of the bicovariant bimodules,
the conditions {\bf B}1 and {\bf B}2 are satisfied.

We also require that the BRST transformation is given by eqs.(\ref{bx3}) and
(\ref{bx4}).
As in the case of the bicovariant differential calculus, the commutation
relation (\ref{c9}) implies that we can write eqs.(\ref{bx3}) and (\ref{bx4})
as the commutator with $C^0$ :

\proposition{4}
\be%
\delta_B C^I=
{ig\over\omega}[C^0,C^I]_+\quad.
\label{c16}
\ee%

\medskip

It is important that the BRST transformation can be represented by the
commutator, since it guarantees the Leibniz rule for the BRST operator
(Cond.{\bf A} 1a).  The nilpotency of the BRST transformation (Cond.{\bf A}1b)
also holds on the algebra of the ghosts due to \eq{c13}.

The covariance of the $\brs$ (Cond.{\bf A}5a) is also clear since $C^0$ is
the invariant element and thus
\be%
\col(\brs \rho)
 ={1 \over \omega}\col([C^0,\rho])
 =(id\otimes \brs)\col(\rho)\quad.
 \label{c23}
\ee%
In order to impose the hermiticity of the ghost field, the commutation relation
(\ref{c02}) must be consistent with the algebra of the ghost fields defined
above.  This means that the $*$-operation which is antimultiplicative in the
graded sense, i.e.,
\be%
(C^IC^J)^*=-(C^J)^*(C^I)^*
\label{c17}
\ee%
is a covariant inner involution of the algebra.
This property is also clear since the algebra of the ghost is the same as the
algebra of the right invariant basis $\theta^I$.  In the present case, it is
straightforward to prove these properties explicitly:

{}From the definition in eqs.(\ref{c02}) and (\ref{c05}), we find
\bea%
(C^-)^* &=&qC^+\label{c18}\quad,\\
(C^+)^* &=&q^{-1}C^-\label{c19}\quad,\\
(C^3)^* &=&C^3\label{c20}\quad,\\
(C^0)^* &=&C^0\quad.\label{c21}
\eea%
By using these relation we see easily
that the relations (\ref{c3})-(\ref{c13}) are
invariant under the $*$-conjugation.
The last equation together with \eq{c16} implies also the condition {\bf A}4a.

\subsection{Algebra of Matter and Ghost Field}

Next we add the matter fields consistently to the above algebra of the ghost
fields.
The commutation relation among the ghost and matter fields
can be derived as follows:

First of all, the requirement of the covariance implies that the commutation
relation is written in terms of linear combinations
 of the possible $\R$ matrix.
Its consistency with the ghost algebra as well as the matter algebra implies
that the matrix giving the commutation relation among
the ghost and matter fields must
satisfy the Yang-Baxter like equations and it excludes the linear combination
of the $\R$ matrix like in the case of the other comodule algebra (see for
example ref.[CSW]).

Therefore, for example if the matter $\Psi$ is bosonic and of the fundamental
representation $\Psi^i$ as given in the appendix C, then we can take the
ansatz:
\be%
 \Psi^k C^{i_1i_2}
 =-\alpha\R^{\pm}{}^{ki_1}_{j_1s}\R^{\pm}{}^{ si_2}_{j_2l}C^{j_1j_2}\Psi^l
 \quad,
 \label{d2}
\ee%
where $\alpha$ is a constant which is $1$ in the limit $q\rightarrow1$.  The
$\pm$ attached to the
 $\R$ matrix means that any combination is allowed and thus
we have four possibilities.
For the general representation we can write the ansatz as follows:
\be%
\Psi C^{i_1i_2}=
\pm\alpha C^{j_1j_2}\big(\Psi*(L^{i_2}_{\sigma
j_2}*L^{i_1}_{\sigma'j_1})\circ\anti\big)\quad,
\label{d1}
\ee%
where we take the $+$ sign for
the bosonic and the $-$ for the fermionic matter. The
suffices $\sigma,\sigma'=\pm$, and $L^i_{\sigma j}$ is the corresponding
functional defined in \eq{a3}. It is easy to see that the relation (\ref{d1})
gives the commutation relation (\ref{d2}) for the boson of the fundamental
representation $\Psi^i$.
Using this ansatz, we can prove the following statement.

\proposition{5}

{\sl
The commutation relation among the ghost and matter is given by
\be%
\Psi C^I=\pm C^J(\Psi*{\bf L}^I_J)\quad,
\label{dx}
\ee%
where the $+$($-$) sign is taken for the bosonic(fermionic) matter.
}

\proof
First we impose the consistency with $*$-conjugation (Cond.{\bf E}3).
The hermiticity of $C^i_j$ in \eq{c02} can be rewritten using the convention
(\ref{c03}) as
\be%
(C^{ij})^*=C^{kl}\epsilon_{kj}\epsilon_{li}\quad.
\label{d5}
\ee%

Then the $*$-conjugation
 of the above ansatz (\ref{d1}) with hermiticity (\ref{d5})
gives the commutation relation between the $\Psi^*$ and $C^{ij}$ as
\bea%
( C^{k_1k_2})\Psi^*
 &=&\pm\alpha^*
 \big(\Psi^**\epsilon^{i_2k_1}\epsilon^{i_1k_2}
 \big(\Psi*(L^{i_2}_{\sigma j_2}*L^{i_1}_{\sigma'j_1})\circ\anti\big)^*
 \circ\anti\epsilon_{l_1j_2}\epsilon_{l_2j_1}\big)C^{l_1l_2}\quad,\nonumber\\
 &=&\pm\alpha^* \big(\Psi^**\epsilon^{i_2k_1}\epsilon^{i_1k_2}
 (\anti^{-1}(L^{j_1}_{\bar{\sigma}'i_1})
 *\anti^{-1}(L^{j_2}_{\bar{\sigma}i_2}))\epsilon_{l_1j_2}\epsilon_{l_2j_1}\big)
 C^{l_1l_2}\quad,\nonumber\\
 &=&\pm\alpha^* \big(\Psi^**L^{k_2}_{\bar{\sigma}'l_2}
 *L^{k_1}_{\bar{\sigma}l_1}\big)C^{l_1l_2}\quad,
 \label{d7}
\eea%
where if $\sigma=\pm$ then $\bar\sigma=\mp$ and the same rule for
$\bar{\sigma}'$.

The above equation is equivalent to
\be%
\Psi^* C^{i_1i_2}=\pm\alpha^{*-1} C^{j_1j_2}
\big(\Psi*(L^{i_2}_{\bar{\sigma}' j_2}
*L^{i_1}_{\bar{\sigma}j_1})\circ\anti\big)\quad.\label{d8}
\ee%
The condition {\bf E}3 implies that \eq{d8} is equivalent to the relation
(\ref{d1}) for the matter $\Psi^*$. Thus we get that
\be%
L^i_{\sigma' j}=L^i_{\bar{\sigma}j}\and \alpha\alpha^*=1\quad.
\label{d8.1}
\ee%

To fix the choice of the $\sigma$ we check the consistency with the BRST
transformation.
For the BRST transformation of the matter we use the form given in \eq{bx5} of
Cond.{\bf C}.
For this part of the proof let us write the index of the matter field
explicitly as $\Psi^a$ where $a$ runs over the corresponding representation.
Taking the BRST transform of both sides of the relation, we get
\be%
\brs(l.h.s.)
={ig\over\omega}\Big\{C^0\Psi^a C^{i_1i_2}-C^J(\Psi^a
*{\bf L}^0_J)C^{i_1i_2}\pm\Psi^a C^0C^{i_1i_2}\pm\Psi^a C^{i_1i_2}C^0\Big\}
\quad,
\label{z1}
\ee%
\be%
\brs(r.h.s.)=\pm\alpha {ig\over\omega}
\Big\{C^0C^{j_1j_2}\big(\Psi^a*{\bf L}^{i_1i_2}_{\sigma j_1j_2}\big)+
C^{j_1j_2}C^J\big(\Psi^a*{\bf L}^{i_1i_2}_{\sigma j_1j_2}*{\bf L}^0_J\big)
\Big\}
\quad,
\label{z2}
\ee%
where ${\bf L}^{i_1i_2}_{\sigma j_1j_2}
=(L^{i_2}_{\sigma j_2}*L^{i_1}_{\bar{\sigma} j_1})\circ\anti$.

Using the ansatz (\ref{d1}) again for the $\brs(l.h.s.)$, we can see the
first terms of eqs.(\ref{z1}) and (\ref{z2}) are equivalent and thus we get the
condition from the equivalence of eqs.(\ref{z1}) and (\ref{z2}):
\be%
C^JC^{K}[(\Psi^a*{\bf L}^0_J*{\bf L}^{I}_{\sigma K})-\alpha
(\Psi^a*{\bf L}^0_{\sigma J}*{\bf L}^I_{\sigma K})
-\alpha (\Psi^a*{\bf L}^I_{\sigma J}*{\bf L}^0_{\sigma K})]
=-C^{J}C^K\big(\Psi^a*{\bf L}^{I}_{\sigma J}*{\bf L}^0_K\big)
\quad.
\label{d20}
\ee%

Multiplying the matrix $\R^{Lb}_{aI}$
\be%
\R^{Lb}_{aI}=\R^{\bar{\sigma}}{}^{l_1a'}_{ai_1}\R^{\sigma}{}^{l_2b}_{a'i_2}
\quad,
\label{d21}
\ee%
where $\R^{\sigma}{}^{ia}_{bj}$ is the $\R$ matrix of the fundamental
representation (suffix i,j) and the representation of the matter field
(suffix a,b).
Taking the summation over $a$ and $I$ we get
\be%
C^K(C^{I}*{\bf L}^J_{\bar\sigma K})[(\Psi^a*{\bf L}^0_J)
-\alpha (\Psi^a*{\bf L}^0_{\sigma J})]
=C^{I}C^K[\alpha (\Psi^a*{\bf L}^0_{\sigma K})-(\Psi^a*{\bf L}^0_K)]
\quad.
\label{d22}
\ee%
Since $(\Psi*{\bf L}^0_{+J})$ and $(\Psi*{\bf L}^0_{-J})=(\Psi*{\bf L}^0_{J})$
terms are independent, we conclude that $\sigma=-$, i.e.,
\be%
{\bf L}^I_{\sigma J}={\bf L}^I_{-J}={\bf L}^I_{J}
\label{d23}
\ee%
and
\be%
\alpha=1
\quad.
\label{d24}
\ee%

\qed

We can deduce the following commutation relations immediately

\corollary{1}
\bea%
d\Psi C^I &=&\mp C^J(d\Psi*{\bf L}^I_J)\quad,\label{g1}\\
\Psi dC^I &=&dC^J(\Psi*{\bf L}^I_J)\quad,\label{g4}\\
d\Psi dC^I &=&dC^J(d\Psi*{\bf L}^I_J)\quad.\label{d27}
\eea%
\proof
Applying the derivative $d$ on \eq{dx} and comparing the term proportional to
$d\Psi$ and $dC^I$ we get eqs.(\ref{g1}) and (\ref{g4}). Then taking again the
derivative of \eq{g1} or \eq{g4} we get \eq{d27}.

\qed

By using the definition of the operator $\chi_I$, the commutation relation in
(\ref{dx}) implies that we can write the BRST transformation of the matter
field in \eq{bx5} as the commutator:

\proposition{6}

{\sl
\be%
\delta_B\Psi={ig\over\omega}[C^0,\Psi]_\mp
\quad,
\label{d3}
\ee%
where we take the commutator $[\cdot,\cdot]_-$ for the bosonic matter and
the anticommutator for fermionic matter.
}

\proof
\be%
{ig\over\omega}[C^0,\Psi]_\mp
={ig\over\omega}(C^0\Psi\mp \Psi C^0)
={ig\over\omega}(C^0\Psi-C^J\Psi*L^0_J)={ig\over\omega}C^{ij}\Psi
*(\e\sigma^0_{ij}-\sigma^0_{kl}{\bf L}^{kl}_{ij}
\quad.
\label{d3.1}
\ee%
Using the definition of $\chi_I$ in \eq{a7} we have \eq{d3}.

\qed

Once the BRST transformation is written in the form of a commutator with
$C^0$, the Leibniz rule (Cond. {\bf A}1a) follows immediately. Conds.{\bf A}4a,
5a are also straightforward.
The proof of the nilpotency (Cond.{\bf A}1b) is
\be%
\delta_B^2\Psi=\delta_B(C^0\Psi-\Psi C^0)=-C^0[C^0\Psi-\Psi C^0]-[C^0\Psi-\Psi
C^0]C^0=0
\quad.
\label{d4}
\ee%

We also show the consistency of \eq{dx} with the commutation relation of the
ghosts and of the matter (Cond.{\bf E}2), although we required it when we
derived the ansatz (\ref{d1}):

In general
the commutation relation of the matter is given by multiplying the projection
operator on the product of two matter fields.
 Thus it is sufficient to prove
that any projection ${\cal P}^{l_1..l_p}_{i_1..i_p}$ of a tensor
$\Psi^{i_1...i_p}$ commutes with the relation (\ref{dx}).  Since any
covariant projection operator is  defined by a combination of the $\R$
matrix, we have
\bea%
{\cal P}^{l_1..l_p}_{i_1..i_p}\Psi^{i_1..i_p}C^I
&=&\pm {\cal P}^{l_1..l_p}_{i_1..i_p}C^J{\bf L}^I_J(M^{i_1}_{j_1}...
M^{i_p}_{j_p})\Psi^{j_1..j_p}C^I\quad,
\nonumber\\
&=&\pm C^J{\bf L}^I_J(M^{l_1}_{l_1}...M^{i_p}_{j_p})
{\cal P}^{i_1..i_p}_{j_1..j_p}\Psi^{j_1..j_p}C^I
\quad.
 \label{d9.1}
\eea%
This proves the consistency with the relation of the matter fields.

The  consistency of \eq{dx} with the relation of the ghost in eqs.(\ref{c1})
and (\ref{c2}) can be proven as follows:
\be%
\Psi C^IC^K=C^JC^L(\Psi*{\bf L}^I_J*{\bf L}^K_L)
\quad.
\label{d9}
\ee%

Thus the sufficient condition is that
\be%
({\cal P}_r,{\cal P}_{r'})^{IJ}_{KL}{\bf L}^K_S*{\bf L}^L_T
={\bf L}^I_K*{\bf L}^J_L({\cal P}_r,{\cal P}_{r'})^{KL}_{ST}
\quad.
\label{d10}
\ee%
This is equivalent to the condition
\be%
({\R}^{s},{\R}^{s'})^{IJ}_{KL}{\bf L}^K_S*{\bf L}^L_T
={\bf L}^I_K*{\bf L}^J_L({\R}^{s},{\R}^{s'})^{KL}_{ST}
\label{d11}
\ee%
for all combination of $s$ and $s'$ where $s,s'=+,-$, since each projector
${\cal P}_r$ can be represented by the linear combination of $\R^\pm$.  Using
the relation among $L_+$ and $L_-$ and the explicit form of $(\R^s,\R^{s'})$
given in the appendix B, it is easy to see that \eq{d11} is satisfied.

Using the similar method we can also derive the commutation relation of
$(\{dC^I\},\{C^I\})$:

\proposition{7}

{\sl
The commutation relation among the ghosts and their derivatives is given by
\be%
dC^IC^J=C^K(dC^I*{\bf L}^J_K)
\quad.
\label{d12}
\ee%
}

\proof
The first half of the proof of the proposition {5} concerning the consistency
with
the $*$-operation is also applicable here and we can set the relation as
\be%
dC^IC^J=\alpha C^K(dC^I*{\bf L}^J_{\sigma K})
\label{d13}
\ee%
where $\alpha^*\alpha=1$.  The hermiticity of the $dC^i_j$
can be simply imposed and does not require new relations.

Then we check the consistency with the BRST transformation.
Taking the derivative of \eq{c16} we get
\be%
\delta_B dC^I=
{-ig\over\omega}\big\{[dC^0,C^I]_-+[dC^I,C^0]_-\big\}
\quad.
\label{d14}
\ee%
Using the $\R$ matrix representation of \eq{d13} we can see that if
$\sigma=-$($\sigma=+$) then  $dC^0$ ($C^0$) is commuting with all elements of
$\{C^I\}$ ($\{dC^I\}$, resp.). Thus we get

\parbox{5cm}{
\begin{eqnarray*}
\qquad\qquad\delta_B dC^I&=&\Bigg\{
\end{eqnarray*}}
 \parbox{10cm}
{ \bea
-{ig\over\omega}[dC^I,C^0]_-\qquad&\for&
\sigma=-\quad,\hfill\qquad\qquad\label{d15}\\
 -{ig\over\omega}[dC^0,C^I]_-\qquad&\for& \sigma=+\quad.\hfill\qquad\qquad
 \label{d16}
\eea}

Now we see that if $\sigma=-$ then the
 BRST transform of $dC^I$ is the same as the
matter and thus the consistency follows from the proposition {5}.
On the other hand, if $\sigma=+$ we can show that
 it contradicts with the nilpotency
of the BRST operation as follows.  Taking the BRST transform of the $r.h.s.$ of
\eq{d16} we get
\be%
\brs(r.h.s.) =dC^0\brs C^I-dC^J(\brs C^I*{\bf L_+}^I_J)
\quad.
\label{d16.1}
\ee%
The above two terms do not cancel and thus $\brs^2$ on $dC^I$ is not zero with
the choice $\sigma=+$.
Finally, to get the Leibniz rule of the BRST operation on the product of $dC^I$
we must set $\alpha=1$.
Thus we get \eq{d12}.

\qed

The consistency of the relation (\ref{d12}) with the algebra of the ghosts is
as follows.  We use the projector expansion of the matrix
$(\R^{-1},\R)^{IJ}_{KL}$ in \eq{h2} of the appendix B.
Acting with $({\cal P}_S,{\cal P}_S)$ and $({\cal P}_A,{\cal P}_A)$ on both
sides of \eq{d12} we get
\bea%
({\cal P}_S,{\cal P}_S)^{IJ}_{KL}(dC^K\,C^L-C^K\,dC^L) &=&0\quad, \label{h3}\\
({\cal P}_A,{\cal P}_A)^{IJ}_{KL}(dC^K\,C^L-C^K\,dC^L) &=&0\quad. \label{h4}
\eea%

These relations show the consistency of the relation (\ref{d12}) with the
eqs.(\ref{c1}) and (\ref{c2}) since eqs.(\ref{h3}) and (\ref{h4}) are the
derivatives of eqs.(\ref{c1}) and (\ref{c2}), respectively. Furthermore
 eqs.(\ref{cc7}) and (\ref{cc8}) follow immediately .

\corollary{2}

{\sl
 $dC^I$ are $q$-commuting  and their commutation relation is defined by
 eqs.(\ref{cc7}) and (\ref{cc8}).
}

 \proof
Taking the derivative of \eq{d12} we get
\be%
dC^IdC^J=(\R^{-},\R)^{IJ}_{KL}dC^K dC^L
\quad,
\label{h5}
\ee%
where we have evaluated the functional ${\bf L}^I_J$. Using the projector
expansion (\ref{h2})
in the appendix B, we get the eqs.(\ref{cc7}) and (\ref{cc8}).

\qed

\subsection{Gauge Field}

The BRST transformation of the gauge field can be derived by the usual logic
used in the nondeformed gauge theory.
The derivative of the field is not covariant under the BRST transformation. Its
transformation is
\bea%
\brs d\Psi
&=&-d\brs\Psi\quad,\nonumber\\
&=&-d[C^{I}(\Psi*\chi_I)]\quad,\nonumber\\
&=&C^{I}(d\Psi*\chi_I)-(dC^I)(\Psi*\chi_I)\quad.\label{e2}
\eea%

We define the covariant derivative $\nabla$ by introducing the gauge field
$A^I$ as \eq{bx6} and we require the covariance under the BRST transformation
(\ref{bx7}) which can be rewritten as
\bea%
\brs \nabla\Psi&=&C^I(\nabla\Psi*\chi_I)\quad,
 \nonumber\\
 &=&C^I(d\Psi*\chi_I)+C^IA^J(\Psi*\chi_I*\chi_J)\quad.\label{e5}
\eea%

On the other hand taking the BRST transformation of the
r.h.s. of \eq{bx6} we get
\bea%
\brs \nabla\Psi
&=&-d\brs\Psi+(\brs A^I)(\Psi*\chi_I)-A^I\brs(\Psi*\chi_I)\quad,\nonumber\\
 &=&-(dC^I)\Psi*\chi_I+C^I(d\Psi*\chi_I)+(\brs
 A^I)(\Psi*\chi_I)-A^IC^J(\Psi*\chi_I*\chi_J)\quad.\label{e6}
\eea%

The BRST transformation of the gauge field can be defined by requiring the
equivalence of the eqs.(\ref{e5}) and (\ref{e6}).
Thus we get
\be%
(\brs A^I)(\Psi*\chi_I)
=(dC^I)(\Psi*\chi_I)+(A^IC^J+C^IA^J)(\Psi*\chi_I*\chi_J)
\quad.
\label{e7}
\ee%

In order to separate the fields from the generators $\chi_I$ of the dual
algebra we need to reduce the $(\chi_I*\chi_J)$ in the second term into a term
 linear in the generators $\chi_I$.  In the usual non-deformed case, we use the
commutator for this purpose.  The corresponding relation in the $q$-deformed
case is \eq{a11}.  Therefore, in our case $A^IC^J+C^IA^J$ need to create the
projector ${\bf P}_{Ad}$ to apply \eq{a11}.

Using the previous result we can prove that

\proposition{8}

{\sl

The commutation relation among the ghost and gauge fields is given by
\be%
A^IC^J =-C^K(A^I*{\bf L}^J_K)
\quad.
\label{e8}
\ee%
With the above relation we can separate the algebra of $\chi_I$
and the BRST transformation of the gauge fields. Then,
the BRST transformation of the gauge fields is given by
\bea%
\brs A^0 &=&dC^0\quad,\label{e8.1}\\
\brs A^a&=&dC^a-ig(\omega C^0A^a+f^a_{bc}C^bA^c)\quad.
\label{e8.2}
\eea%
}

\proof
We can apply the same argument of the proof of the proposition {5} concerning
the consistency with the $*$-operation and we can set
\be%
A^IC^J =-\alpha C^K(A^I*{\bf L}^J_{\sigma K})
\quad.
\label{e8.3}
\ee%
To fix the choice of $\sigma$ we apply the BRST transformation of \eq{e8.3}.
The hermiticity of $A^I$ can be imposed simply and does not require any new
condition.
We know from \eq{e7} that
\be%
\brs A^I=dC^I+\{other\ terms\ independent\ of \ dC^I\}
\label{ee8.3}
\ee%
Thus comparing the $dC^I$ dependent term of the
BRST transform of \eq{e8.3} and
\eq{d12} we conclude that the relation of the ghost and gauge
fields is \eq{e8}.

Using the definition of the ${\bf L}$, the relation (\ref{e8}) in terms
of the $\R$ matrix is
\be%
A^IC^J =-(\R^{-1},\R)^{IJ}_{KL}C^KA^L
\quad.
\label{e8'}
\ee%
Substituting this into \eq{e7}, we get
\bea%
\brs A^I(\Psi*\chi_I)
&=&(dC^I)(\Psi*\chi_I)+C^KA^L(({\bf1},{\bf1})^{IJ}_{KL}-(\R^{-1},\R)^{IJ}_{KL})
(\Psi*\chi_I*\chi_J)
\quad,\nonumber\\
 &=&(dC^0)(\Psi*\chi_0) +[dC^a-
 ig(\omega C^0A^a+f^a_{bc}C^bA^c)](\Psi*\chi_a)
 \quad,
 \label{e15}
\eea%
where we have used \eq{ap1} in the appendix B.
Thus we get eqs.(\ref{e8.1}) and (\ref{e8.2})

\qed

Then applying the derivative $d$ on \eq{e8}, we get

\corollary{3}
\bea%
dA^IC^J &=&C^K(dA^I*{\bf L}^J_K)\quad,\label{g9}\\
A^IdC^J &=&dC^K(A^I*{\bf L}^J_K)\quad,\label{g10}\\
dA^IdC^J &=&dC^K(dA^I*{\bf L}^J_K)\quad.\label{g10.1}
\eea%

\proof
Applying $d$ on \eq{e8} and comparing the terms proportional to
$dA^I$ and $dC^I$ we get eqs.(\ref{g9}) and (\ref{g10}). Applying $d$ on
\eq{g10} we get \eq{g10.1}.
\qed

To prove the nilpotency of the BRST transformation, we show that the BRST
transformation of the gauge field can be also written by using the commutator
with $C^0$.

\proposition{9}

{\sl
The BRST transformation of the gauge field can be expressed as
\be%
\brs A^I=dC^I+
{ig\over\omega}[A^I,C^0]_+
\label{e9.1}
\ee%
and the BRST operation is nilpotent.
}

\proof
Using the decomposition of the $( \R^-,\R)$ given in the appendix B,
the relation of
$A^I$ and $C^I$ can be written in components as
\bea%
A^0C^0 &=&-C^0A^0\quad,\label{e9}\\
A^0C^a &=&-C^aA^0\quad,\label{e10}\\
A^aC^0 &=&-(\omega^2+1)C^0A^a-\omega f^a_{bc}C^bA^c\quad,\label{e11}\\
A^aC^b
&=&-C^aA^b-f^{ab}_{a'}(\omega C^0A^{a'}+ f^{a'}_{cd}C^cA^d)\quad.\label{e13}
\eea%
The third relation can be written as
\be%
{1 \over \omega }[A^a,C^0]_+=-(\omega C^0A^a+f^a_{bc}C^bA^c)\quad.
\label{e14}
\ee%
Comparing the above eqs.(\ref{e9}) and (\ref{e14}) with eqs.(\ref{e8.1}) and
(\ref{e8.2}), we get the formula given in \eq{e9.1}.

The nilpotency of the BRST operation on $A^0$ is apparent. The BRST operation
on the gauge field $A^a$ can be proven as
\bea%
\brs(\brs A^a)
&=&-d\brs C^a+{ig \over \omega }[\brs A^a,C^0]_-\quad,\nonumber\\
&=&{ig \over \omega }(d[C^0,C^a]_+-[dC^a,C^0]_-
+{ig \over \omega}[[A^a,C^0]_+,C^0]_-)\quad,\nonumber\\
&=&-{ig \over \omega }([dC^0,C^a]_-)=0
\quad.
\label{e14.1}
\eea%
\qed

Now using the BRST transformation of the gauge field given in the
proposition {9} it follows:

\proposition{10}

{\sl
\be%
\Psi A^I=\pm A^I(\Psi*{\bf L}^I_J)
\label{g5}
\ee%
where we take the $+$ sign for bosons and the $-$ for fermions.
}

\proof
Applying again the same argument concerning the $*$-operation in the
proposition {5} replacing $C^I$ with $A^I$, since $A^I$ is also given by a
hermitian matrix,
we can set the ansatz as:
\be%
\Psi A^I =\pm\alpha A^J(\Psi*{\bf L}^I_{\sigma J})
\label{g6}
\ee%
To fix the choice of $\sigma$ we apply the BRST operation on \eq{g6}.
Comparing  the term proportional to $dC^I$ of the result with \eq{g4}, we
conclude that $\sigma=-$ and $\alpha=1$.

\qed

\corollary{4}
\bea%
d\Psi A^I&=&\mp A^I(d\Psi*{\bf L}^I_J)\quad,\label{g6.1}\\
\Psi dA^I&=& dA^I(\Psi*{\bf L}^I_J)\quad,\label{g6.2}\\
d\Psi dA^I&=& dA^I(d\Psi*{\bf L}^I_J)\quad.\label{g6.3}
\eea%
\proof
Apply the derivative $d$ on \eq{g6}, then eqs.(\ref{g6.1}) and (\ref{g6.2})
follow using the independence of the terms proportional to $dA^I$ and $d\Psi$.
Applying again the derivative on \eq{g6.1} we get \eq{g6.3}.

\qed

As a result of proposition {10}, we can also prove the following relations

\corollary{5}
\be
\nabla\Psi=d\Psi+{ig\over\omega}[A^0,\Psi]_{\mp}
\quad.
\label{g6.4}
\ee
\proof For the derivative part it is trivial. For the commutator part, replace
the $C^0$ in \eq{d3.1} with $A^0$ then we see that the second term in \eq{g6.4}
  is equivalent to $A^I(\Psi*\chi_I)$.
\qed

{}From this representation of the covariant derivative,
the Leibniz rule and the conditions \eq{bx8} follow immediately
since $A^{0*}=A^0$.

{}From the above results concerning the covariant derivative we can prove the
following relation.

\proposition{11}

{\sl
The BRST transformation of the covariant derivative of
the matter is written as the
commutator with $C^0$:
\bea%
\brs\nabla\Psi
&=&{ig\over\omega}[C^0,\nabla\Psi]_\pm
\label{g3}
\eea%
and thus
the nilpotency of the $\brs$ operator on $\nabla\Psi$ follows.
}

\proof
{}From the representation of the covariant derivative given in \eq{g6.4}, it is
easy to show the relation
\be%
\nabla\Psi C^I=\mp C^J(\nabla\Psi*{\bf L}^I_J)
\quad.
\label{g2}
\ee%
We show that each term in \eq{g2} satisfies the above relation separately.
For the first term of \eq{g6.4}, \eq{g2} is clear from \eq{g1}.
Concerning the second term, due to \eq{e9} $A^0$ simply makes the term opposite
statistics and thus the commutation relation is followd from \eq{dx} with the
opposite sign. Thus, we get \eq{g2}. Then, replace the $\Psi$ with $\nabla\Psi$
and use the \eq{g2} instead of \eq{dx} in the proof of the proposition 6.

\qed

\subsection{Field Strength}

The field strength $F^I$ is defined by the square of the covariant derivative.
When we apply it on the matter $\Psi$ we get
\bea%
F^I(\Psi*\chi_I)&\equiv&\nabla(\nabla\Psi)\quad,\nonumber\\
&=&d(\nabla\Psi)+A^I((\nabla\Psi)*\chi_I)\quad,\nonumber\\
&=&(dA^I)(\Psi*\chi_I)-{1 \over q^2+q^{-2}}A^bA^cf^a_{bc}
[(\Psi*\opst*\chi_a)+\omega(\Psi*\chi_0*\chi_a)]
\quad.
\label{g7}
\eea%

Therefore, using \eq{a12} we get the field strength as eqs.(\ref{cc11}) and
(\ref{cc12}) in proposition {3}.  Using the commutation relation of the gauge
field it can also be written as
\be%
F^I =dA^I+{ig \over \omega }[A^0,A^I]_+
\quad.
\label{g8}
\ee%

Using the covariance of the $\nabla^2\Psi$, we can prove the covariance of the
gauge fields and derive the following commutation relations.

\proposition{12}
\be%
F^IC^J=C^K(F^I*{\bf L}^J_K)
\quad.
\label{g13}
\ee%
\proof
It is straightforward to prove it directly using eqs. (\ref{e8}), (\ref{g9})
and (\ref{g8}).  We can also derive the relation as follows:
Since $\nabla^2\Psi$ is covariant we can apply the proof of the
proposition {5} and thus
\be%
[F^I(\Psi*\chi_I)]C^J=
C^K\big([F^I(\Psi*\chi_I)]*{\bf L}^J_K\big)=C^K(F^I*{\bf L}^J_{K'})
(\Psi*\chi_I*{\bf L}^{K'}_K)
\quad,
\label{g11}
\ee%
where we have used the coproduct of the functional ${\bf L}^I_J$.
On the other hand the $r.h.s.$ can be written as
\be%
r.h.s.=F^IC^K(\Psi*\chi_I*{\bf L}^J_K)
\quad.
\label{g12}
\ee%
{}From the equivalence of these two equations we get \eq{g13}.

\qed

\proposition{13}

{
\sl
The field strength defined in \eq{g8} is transformed under the BRST
transformation as \eq{cc13} in proposition {3} and equivalently by the
commutator as
\be%
\brs F^I={ig\over \omega}[C^0, F^I]_-
\quad.
\label{g14}
\ee%
}

\proof
The BRST transformation of the field strength is
\bea%
\brs F^a
 &=&-d\brs A^a+{ig\over \omega}([\brs A^0,A^a]_-+[\brs A^a,A^0]_-)\quad,
 \nonumber\\
 &=&{ig \over \omega}(-d[A^a,C^0]_++[dC^0,A^a]_-+[dC^a,A^0]_-
 +{ig\over \omega}[[A^a,C^0]_+,A^0]_-)\quad,
 \nonumber\\
 &=&{ig \over \omega}([C^0,F^a]_-+[dC^a,A^0]_-)
 \quad.
 \label{g15}
\eea%

Using the commutation relation (\ref{g10}) we get \eq{g14}.
Then using \eq{g13},
we get  \eq{cc13}.

\qed
\proposition{14}

{\sl
The commutation relation of the fields $A^I$ and their derivatives $dA^I$
is given by
\be%
dA^IA^J=A^K(dA^I*{\bf L}^J_K)
\quad.
\label{g20}
\ee%
}

\proof
  Using the same argument concerning the $*$-conjugation in the proof of the
  proposition {5}, they must satisfy
\be%
dA^IA^J=\alpha A^K(dA^I*{\bf L}^J_{\sigma K})
\quad.
\label{g221}
\ee%
Take the BRST transformation of the \eq{g221} and compare the term proportional
to the derivative of the ghost $dC^I$. Then, we find
\be%
dA^IdC^J=\alpha dC^K(dA^I*{\bf L}^J_{\sigma K})
\quad.
\label{g222}
\ee%
Since this must be equivalent to \eq{g10.1},
we conclude $\alpha=1$ and $\sigma=-$.
The consistency of the \eq{g20} with the commutation relation among the $A^I$
can be shown analogously to the ghost case by replacing $C^I$ and $dC^I$ in
eqs.(\ref{h3}) and (\ref{h4}).

\qed

\corollary{6}

{\sl

 $dA^I$ are $q$-commuting  and their commutation relation is defined by
 eqs.(\ref{cc5}) and (\ref{cc6}).
}

\proof
Taking the derivative of \eq{g20} and use the same formula used in the proof of
the corollary {2}.

\qed

Finally we give the formula which is the simpler representation of the Bianchi
identity in  \eq{cc14} of proposition {3}.

\proposition{15}

{\sl

The field strength $F^I$ satisfies the following relation
\be%
dF^I={ig\over \omega}[F^I,A^0]_-
\quad.
\label{g24}
\ee%
}

\proof
Taking the derivative of the field strength $F^I$ given in \eq{g8}, we get
\be%
dF^I
={ig \over \omega }\Big\{[dA^0,A^I]_-+[dA^I,A^0]_-\Big\}={ig \over \omega }
[dA^I,A^0]_-
\quad.
\label{g25}
\ee%
On the other hand using the relation $A^0A^0=0$ we get
\be%
[F^I,A^0]_-=
[dA^I,A^0]_-+{ig \over \omega }[[A^0,A^I]_+,A^0]_-
=[dA^I,A^0]_-
\quad.
\label{g26}
\ee%
Thus we get \eq{g24}.

\qed

\subsection{Proof of Propositions 1-3 and Summary}

We complete here the proof of the propositions 1, 2, and 3:

\medskip
\noindent{Proof of proposition 1}

All commuation relations among the fields given in the
propositions 5, 7, 8, 10, 14 and the corollaries 1, 3, 4,
 can be summarized as in eqs.(\ref{cc3}) and (\ref{cc4}).
The corollary 2 and 6 prove (\ref{cc5})-(\ref{cc8}).

\qed

\noindent{Proof of proposition 2}

Substituting the explicit relations (\ref{e9})  and (\ref{e14})
into the \eq{e9.1} in the proposition 9, we get the eqs.(\ref{cc9})
and (\ref{cc10}).  Then the nilpotency follows from the proposition 9.

\qed

\noindent{Proof of proposition 3}

As we mentioned \eq{g7} gives the definition of the field strength as
eqs.(\ref{cc11}) and (\ref{cc12}).  Using the commutation relation given in
the proposition 12, the BRST transformation of $F^I$ given in proposition 13
can be written as \eq{cc13}.
The Bianchi identity in proposition 3 can be derived as follows:
As for the $F^0$ it is trivial.  To prove \eq{cc14}, we substitute
 \eq{cc11} into the $r.h.s$ of \eq{g24}.
Since the commutation relation of $A^I$ and $C^I$ are given by the same formula
(\ref{c1})-(\ref{cc2}), we have the relation for the $[A^0,A^a]_+$
corresponding to \eq{c9}.  Therefore we get
\bea
[F^a,A^0]_-
&=&{-\omega\over q^2+q^{-2}}f^a_{bc}d(A^bA^c)
-{ig\over q^2+q^{-2}}f^a_{bc}[A^bA^c,A^0]\nonumber\\
&=&{-\omega\over q^2+q^{-2}}\Big\{f^a_{bc}
(dA^bA^c-A^bdA^c)+{ig\over\omega}f^a_{bc}(-[A^b,A^0]A^c+A^b[A^c,A^0])\Big\}
\eea
Using \eq{g8}, we get \eq{cc14}.

\qed

The propositions 4, 6, 11, 13 can be summarized as: the BRST transformation of
the fields except gauge fields can be represented by the commutator with $C^0$.
Concerning the gauge field, the proposition 9 shows that the homogeneous term
of the BRST transformation is again a commutator with $C^0$.

The above property is very important in the present construction of
the BRST algebra, since due to this property, the Leibniz rule of the operator
$\brs$ is satisfied in rather trivial way.  The nilpotency becomes
 also apparent since $(C^0)^2=0$.

\section{Discussion}

We have constructed the $q$-deformed BRST algebra which corresponds to the $q$
deformation of the algebra of the ghost, gauge and matter fields on one
spacetime point.  To obtain the $q$-deformation of the BRST formulation of the
gauge field theory, we have to take the structure of the base manifold into
consideration.  Using the result here,
 one may take the base manifold as a usual spacetime but a
 more
interesting possibility is the one when the base manifold is also described by
the non-commutative function algebra.
In both cases, we have to reconsider the meaning
of the usual quantization so that it fit to the pure algebraic formulation.

In the latter case, we also have to
 consider the new theory of the gravity based on the noncommutative geometry.
  Then we can ask the interesting question whether
the new deformation parameter may play the role of the cut off of the quantum
 gravity.
Our result here may also shed some light on that direction, since the local
Lorentz symmetry can be treated as usual gauge symmetry. It is known that
 there is a quantum
 deformation of the Lorentz group [CSSW,PW] and thus one
may write down the $q$-deformed BRST algebra using the result of the
differential calculus on the quantum Lorentz group [CDSWZ,CW,SWW].

In the construction presented here, we used the singlet component in the
hermitian matrix.  Especially the proof of the nilpotency and Leibniz rules
became simple, since all action of the $\brs$ operator
is represented by the
commutator with $C^0$.  Nevertheless, it is very interesting if there is a
$q$-deformed BRST algebra without singlet component.

Concerning the ghost algebra, it is possible to obtain the algebra without the
singlet component.  First of all, the Maurer-Cartan equation does not contain
 the singlet component and thus the BRST transformation of the ghosts does not
either. The algebra of $\chi_I$
contains the singlet component but it appears in
the functional $\opst$
and this $\opst$ is central in the algebra, therefore we
can divide it out.  Then we get the algebra
where the commutation relations are given by setting $\opst=i$ in the
eqs.(\ref{a13})-(\ref{a14'}).
With this new operator $\tilde\chi_a=\chi_a*\opst^{-1}$ we can write
$\brs\Psi=C^a(\Psi*\tilde\chi_a)$ and prove that $\brs^2=0$.
In this way we can remove the singlet component in the BRST transformation
 and thus we do not need the
singlet component of the gauge field.
The trouble occurs when we start to consider the commutation relations among
the different types of fields such as gauge and ghost fields without the
singlet component.
We can not use the $\R$ matrix which is defined in the
standard way for the tensor
representation. For example if we simply take
\be%
C^aA^b=\R^{ab}_{cd} A^cC^d
\label{disc1}
\ee%
where $\R^{ab}_{cd}$ is the $\R$ matrix of adjoint representation.
If we take the $*$-conjugation and using the hermiticity of the fields,
\be%
A^aC^b=\R^{ab}_{cd} A^cC^d\label{disc2}
\ee%
and thus we can not impose the hermiticity condition.

\bigskip
\noindent{\Large\bf Acknowledgement}

The author would lilke to acknowledge  U. Carow-Watamura, M. Hirayama, P.P.
Kulish,  R. Sasaki  for interesting and critical discussions.  He thanks K.
Ueno and T. Masuda for helpful discussions on the mathematical aspects.  He
would also like to acknowledge M. Schmidt for his kind hospitality during the
stay at the Institute for Theoretical Physics of the Heidelberg University.

\startapp

\bigskip

\noindent{\Large\bf Appendix}
\medskip

\sectione{Functionals and Convolution Product}

The quantum matrix is denoted by $M^i_j$ and quantum group relations are
defined by
\bea
\R^{ij}_{k'l'}M^{k'}_kM^{l'}_l &=&M^i_{i'}M^j_{j'}\R^{i'j'}_{kl}\quad,\\
\epsilon_{ij}M^i_kM^j_l&=&\epsilon_{kl}\quad,\\
(M^i_j)^*&=&\anti(M^j_i)\quad.
\eea
There are functionals defined by the following relations
\be%
L^i_{\pm j}(M^k_l)=\R^\pm{}^{ik}_{lj}
\quad.
\label{a3}
\ee%
where the normalization of the $\R$ matrix is
\be%
\R=q^{\half}{\cal P}_S-q^{-3\over2}{\cal P}_A
\label{apr}
\ee%
They satisfy the following relations
\bea
\R^{ij}_{k'l'}L^{l'}_{\pm l}*L^{k'}_{\pm k} &
=&L^j_{\pm j'}*L^i_{\pm i'}\R^{i'j'}_{kl}\quad,\\
\R^{ij}_{k'l'}L^{l'}_{+ l}*L^{k'}_{- k} &
=&L^j_{- j'}*L^i_{+ i'}\R^{i'j'}_{kl}\quad,\\
(L^i_{+j})^*&
=&\anti^{-1}(L^j_{-i})
\quad.
\eea

The convolution product of two functionals $f,g\in{\cal U}_q(su(2))$ is defined
by
\be%
f*g=(f\otimes g)\co
\quad.
\label{ap6}
\ee%
We use the convolution product $*$ between a functional $f$ over the Hopf
algebra, e.g. $f\in {\cal U}_q(su(2))$ and a (left)comodule $\rho$ as (See also
the last section of [CW].)
\be%
(\rho*f)\equiv (f\otimes id)\col(\rho)
\quad.
\label{bb}
\ee%
The $*$-operation is defined as
\be
(\rho*f)^*=(\rho^**f^*\circ\anti)
\quad.
\ee
With these definitions we have the following associativity
\be%
\rho*(f*g)=(\rho*f)*g
\label{ap7}
\ee%
due to the relation $(\co\otimes id)\col=(id\otimes \col)\col$.

\sectione{\bf $\R$ matrix and Projectors}

Given tensors $A^{ij}_{kl}$ and $B^{ij}_{kl}$ we define
\be%
(A,B)^{IJ}_{KL}=(A,B)^{i_1i_2j_1j_2}_{k_1k_2l_1l_2}
=\R^{-}{}^{i_2j_1}_{j_1'i_2'}A^{i_1j_1'}_{k_1l_1'}B^{i_2'j_2}_{k_2'l_2}
\R^{l_1'k_2'}_{k_2l_1}
\quad.
\label{ap5}
\ee%
Applying this notation the pair of the $\R$ matrix, $(\R^{-1},\R)$  can be
expanded by the projection operators as
\be%
(\R^{-1},\R)^{IJ}_{KL}=\big(({\cal P}_S,{\cal P}_S)+({\cal P}_A,{\cal P}_A)
-q^2({\cal P}_A,{\cal P}_S)-q^{-2}({\cal P}_S,{\cal P}_A)\big)^{IJ}_{KL}
\quad.
\label{h2}
\ee%
Note that each term $({\cal P}_r,{\cal P}_{r'})$ is a projection operator.
(see also ref.[CSWW].)
Then using the $\sigma^I$ matrices we can prove the following relations.
\bea%
\lefteqn{\delta^{i_1}_{k_1} \delta^{i_2}_{k_2}
\delta^{j_1}_{l_1} \delta^{j_2}_{l_2}
-
\R^-{}^{i_1j_3}_{k_1i_3}
\R^-{}^{i_2j_1}_{j_3i_4}
\R^{i_4j_2}_{j_4l_2}
\R^{i_3j_4}_{k_2l_1}}\nonumber\\
&=&\sigma^{i_1i_2}_a\sigma^a_{k_1k_2}\sigma^{j_1j_2}_0\sigma^0_{l_1l_2}
-\sigma^{i_1i_2}_0\sigma^a_{k_1k_2}\sigma^{j_1j_2}_a\sigma^0_{l_1l_2}
+\sigma^{i_1i_2}_0\sigma^0_{k_1k_2}\sigma^{j_1j_2}_a\sigma^a_{l_1l_2}
-(\omega^2+1)\sigma^{i_1i_2}_a\sigma^0_{k_1k_2}\sigma^{j_1j_2}_0
\sigma^a_{l_1l_2}\nonumber\\
&&-\omega f_a^{bc}\sigma^{i_1i_2}_b\sigma^0_{k_1k_2}\sigma^{j_1j_2}_c
 \sigma^a_{l_1l_2}
 -\omega f^a_{bc}\sigma^{i_1i_2}_a\sigma^b_{k_1k_2}\sigma^{j_1j_2}_0
 \sigma^c_{l_1l_2}
 +(q^2+q^{-2}){\bf P}\!_{Ad\ cd}^{\ ab}
 \sigma^{i_1i_2}_a\sigma^c_{k_1k_2}\sigma^{j_1j_2}_b\sigma^d_{l_1l_2}\ ,
 \nonumber\\
\label{ap1}
\eea%
and
\bea%
\lefteqn{\delta^{i_1}_{k_1} \delta^{i_2}_{k_2}
\delta^{j_1}_{l_1} \delta^{j_2}_{l_2}
-
\R^{i_1j_3}_{k_1i_3}
\R^-{}^{i_2j_1}_{j_3i_4}
\R^-{}^{i_4j_2}_{j_4l_2}
\R^{i_3j_4}_{k_2l_1}}\nonumber\\
&
=&\sigma^{i_1i_2}_a\sigma^a_{k_1k_2}\sigma^{j_1j_2}_0\sigma^0_{l_1l_2}
-\sigma^{i_1i_2}_a\sigma^0_{k_1k_2}\sigma^{j_1j_2}_0\sigma^a_{l_1l_2}
+\sigma^{i_1i_2}_0\sigma^0_{k_1k_2}\sigma^{j_1j_2}_a\sigma^a_{l_1l_2}
-(\omega^2+1)\sigma^{i_1i_2}_0\sigma^a_{k_1k_2}\sigma^{j_1j_2}_a
      \sigma^0_{l_1l_2}\nonumber\\
 &&-\omega f_a^{bc}\sigma^{i_1i_2}_b\sigma^a_{k_1k_2}\sigma^{j_1j_2}_c
 \sigma^0_{l_1l_2}
 -\omega f^a_{bc}\sigma^{i_1i_2}_0\sigma^b_{k_1k_2}\sigma^{j_1j_2}_a
 \sigma^c_{l_1l_2}
 +(q^2+q^{-2}){\bf P}\!_{Ad\ cd}^{\ ab}
 \sigma^{i_1i_2}_a\sigma^c_{k_1k_2}\sigma^{j_1j_2}_b\sigma^d_{l_1l_2} \ ,
\nonumber\\
\label{ap2}
\eea%
Thus the product of the two hermitian fields $A^I$ and $C^J$ can be
expanded as
\bea%
A^KC^L(({\bf1},{\bf1})-(\R,\R^-))^{IJ}_{KL}(\chi_I*\chi_J)
 =ig( -\omega A^0C^a - f^a_{bc}A^bC^c)\chi_a\quad.\label{ap3}
\eea%

\sectione{\bf Algebra of Matter field}

We give here an example of the algebra of matter fields.
In general, once we define the representation of the matter fields,
 we construct
the corresponding $q$-antisymmetrizer and $q$-symmetrizer.  Then if one wants
 to
define the fermions, we impose the $q$-symmetric product to vanish and the
$q$-antisymmetric product to vanish for the bosons.

When the matter is bosonic and of the fundamental representation $\Psi^i$, the
algebra is defined like the differential calculus on the quantum space
[Pusz,WZ](see also [CSW].), and we can set the following commutation relations:
\bea
{\cal P}_A{}^{ij}_{kl}\Psi^k\Psi^l&=&0\quad,
\label{c30}\\
d\Psi^i\Psi^j-q^{{3\over2}}\Psi^k(\Psi^i*L^j_{+k})&=&0\quad,
\label{c31}
\eea
and the conjugated relation for the fields $\Psi^{*i}$ and $d\Psi^{*i}$.
The relation between the $\Psi^i$ and $\Psi^{*i}$ can be defined for example as
\be
\Psi^{*i}\Psi^j=\beta \Psi^k(\Psi^{*i}*L^j_{+k})\quad,\\
\ee
where $\beta$ is a constant which is $1$ in the limit $q\rightarrow1$.
Then, by the consistency with derivative $d$, it is easy to show
the following relations:
\bea
\Psi^{*i}d\Psi^j&=&\beta d\Psi^k(\Psi^{*i}*L^j_{+k})\quad,\\
d\Psi^{*i}d\Psi^j&=&-\beta d\Psi^k(d\Psi^{*i}*L^j_{+k})\quad,\\
{\cal P}_S{}^{ij}_{kl} d\Psi^kd\Psi^l&=&0\quad,\label{c32}
\eea
and their $*$-conjugation.

\bigskip
\noindent{\Large\bf References}

\begin{tabular}{lp{14cm}r}

[Abe]&E. Abe, "Hopf Algebras", Cambridge Tracts in Math., vol.74, Cambridge
Univ. Press, 1980.\\[1ex]

[Are]& I.Ya. Aref'eva, I.V. Volovich,
	Quantum group gauge fields.
	Mod. Phys. Lett. {\bf A6} (1991) 893-907.\\[1ex]

[Ber]&D. Bernard, Prog. Theor. Phys. Suppl. {\bf102} 49-66 (1990).\\[1ex]

[BM] &T. Brzezi\'nski and S. Majid,
	"Quantum group gauge theory on quantum spaces."
	Preprint DAMTP/92-27, 1992; "Quantum Group Gauge
	Theory on Classical Spaces" Preprint DAMTP/92-51.\\[1ex]

[BRST]&C. Becchi, A. Rouet and R. Stora, Ann. Phys. {\bf98} 287-321 (1976);\\
	    &I.V. Tyuitin, Lebedev preprint FIAN 39 (1975), unpublished.\\[1ex]

[Carow]&U. Carow-Watamura, Ph.D. Thesis, "The quantum group symmetry in the
models of elementary particle physics."\\[1ex]

[CDSWZ]&C. Chryssomalakos, B. Drabant, M. Schlieker, W. Weich and B. Zumino,
        Commun. Math. Phys. {\bf147}, 635-646 (1992).\\[1ex]

[Connes]&A. Connes, Non-commutative differential geometry, Technical Report 62,
	IHES, 1986.\\[1ex]

[CSSW]&U. Carow-Watamura, M. Schlieker, M. Scholl and S. Watamura,
	Z.Phys. C -- Particles and Fields {\bf 48}, 159-165 (1990);
	Int.Jour. of Mod.Phys. {\bf A}6, 3081-3108 (1991).\\[1ex]

[CSW]&U. Carow-Watamura, M. Schlieker and S. Watamura,
	Z.Phys. C -- Particles and Fields {\bf49}, 439-446 (1991).\\[1ex]

\end{tabular}

\begin{tabular}{lp{14cm}r}

[CSWW]&U. Carow-Watamura, M. Schlieker, S. Watamura and W. Weich,
       Commun. Math. Physics {\bf 142 }, 605-641 (1991).\\[1ex]

[CW]&U. Carow-Watamura and S. Watamura,
  "Complex Quantum Group, Dual Algebra and Bicovariant Differential Calculus,"
  preprint TU-382(1991) to be published in Commun. Math. Physics.\\[1ex]

[Dri]&V.G. Drinfeld, "Quantum Groups",
	Proceedings of the International
	Congress of Mathematicians, 1986, Vol.{\bf 1}, 798-820.\\[1ex]

[FP] &L.D. Faddeev and V. Popov, Phys. Lett. {\bf25B}, 29-30 (1967).\\[1ex]

[FRT] &L.D. Faddeev, N.Yu. Reshetikhin, and L.A. Takhtajan,
	Quantization of Lie groups and Lie algebras.
	Algebra i Analiz, {\bf1} 178-206 (1989).\\[1ex]

[Hira] &M. Hirayama,
        Gauge Field Theory of the Quantum Group $SU_q(2)$,
	preprint TOYAMA-74 1992.\\[1ex]

[IP] &A.P. Isaev and Z. Popowicz
	Phys. Lett. {\bf281B}(1992)271-278.\\[1ex]

[Jim]&M. Jimbo, Lett. Math. Phys. {\bf 10}, 63-69 (1986).\\[1ex]

[Jur]&B. Jurco, Lett. Math. Phys. {\bf 22}, 177-186 (1991).\\[1ex]

[KO]&T. Kugo and I. Ojima,
  Prog. Theore. Phys. Supple. {\bf 66} 1-130 (1979).\\[1ex]

[Manin]&Yu. I. Manin, "Quantum groups and non-commutative geometry",
Centre de Recherches Mathematiques, Universite de Montreal, (1988).\\[1ex]

[Pusz]&W. Pusz and S.L. Woronowicz, Reports Math. Phys.
27,231-257(1989).\\[1ex]

[PW]&P. Podl\'es and S.L. Woronowicz, Commun. Math. Phys. {\bf130}, 381-431
(1990).\\[1ex]

[Rosso]&M. Rosso, Duke Math. J. {\bf61}, 11-40 (1990).\\[1ex]

[SWW]&X. Sun, S. Wang and K. Wu, "Differential Calculus on Quantum Lorentz
Group," preprint CCAST-92-14(1992).\\[1ex]

[Weich]&W. Weich, Ph.D. Thesis, "Die Quantengruppe $SU_q(2)$ - kovariante
Differentialrechnung und ein quantensysmmetrisches quantenmechanisches
Modell", Karlsruhe University, Norvember 1990.\\[1ex]

[Wor1]&S.L. Woronowicz, Commun. Math. Phys. {\bf 111}, 613-665 (1987).\\[1ex]

[Wor2]&S.L. Woronowicz, Commun. Math. Phys. {\bf 122}, 122-170 (1989).\\[1ex]

[WuZ]&K. Wu and R. Zhang, Commun. Theor. Phys. {\bf17}, 175-182 (1992).\\[1ex]

[WZ] &J. Wess and B. Zumino.
         Covariant differential calculus on the quantum hyperplane.
         Nucl. Phys. B (Proc. Suppl.) {\bf 18 B} 302-312, (1990).\\[1ex]

\end{tabular}

\end{document}